\documentclass[useAMS,usenatbib,twocolumn]{mn2e}

\usepackage{graphicx}

\usepackage{subfigure}
\usepackage{xcolor}
\usepackage{mathtext,bbm,amsmath,amsfonts,amssymb,indentfirst,syntonly,graphicx}
\usepackage{mathtools}
\usepackage{slashbox}
\usepackage{calc}
\usepackage{tikz}

\usepackage[english]{babel}

\title[Constrain IDE models by latest observations]{Constraining interacting dark energy models with latest cosmological observations}
\author[D.-M. Xia and S. Wang]{Dong-Mei Xia$^{1}$\thanks{E-mail: xiadm@cqu.edu.cn} and Sai Wang$^{2}$\thanks{E-mail: physics0911@163.com}\\
$^{1}$College of Power Engineering, Chongqing University, Chongqing, China\\
$^{2}$Department of Physics, The Chinese University of Hong Kong, Shatin, NT, Hong Kong SAR, China}
\begin{document}

\date{Accepted xxxxxx. Received xxxxxx; in original form xxxxxx}

\pagerange{\pageref{firstpage}--\pageref{lastpage}} \pubyear{2016}

\maketitle

\label{firstpage}

\begin{abstract}
The local measurement of $H_0$ is in tension with the prediction of $\Lambda$CDM model based on the Planck data. This tension may imply that dark energy is strengthened in the late-time Universe. We employ the latest cosmological observations on CMB, BAO, LSS, SNe, $H(z)$ and $H_0$ to constrain several interacting dark energy models. Our results show no significant indications for the interaction between dark energy and dark matter. The $H_0$ tension can be moderately alleviated, but not totally released.
\end{abstract}


\section{Introduction}

The Hubble parameter $H$ brings important information of our Universe. It is dynamically determined by the Friedmann equations, and then evolves with cosmological redshift. The evolution of Hubble parameter is closely related with the cosmic inventories, including radiations, baryon, cold dark matter, and dark energy, or even other exotic components in the Universe. Further, it may be impacted by some interactions between these inventories. Thus one can spy upon the evolution of the Universe by measuring the Hubble parameter. Measuring $H_0$ could give a stringent test of the standard cosmological model, or provide evidence for some new physics beyond the standard model.

The Hubble constant $H_0$, today's Hubble parameter with redshift $z=0$, has been precisely measured by many approaches. For instance, the Planck Collaboration \citep{Ade:2015xua} have obtained a severe constraint on $H_0$ by observing the cosmic microwave background (CMB) which is formed in a large redshift $z\simeq1090$. This constraint is given by $H_0=67.27\pm0.66~\rm{km/s/Mpc}$ in the framework of base $\Lambda$CDM model. Here the $1\sigma$ uncertainty has been reduced to a $1\%$ level. With 300 supernovae of type Ia (SNe Ia) at $z<0.15$, recently, the Hubble constant $H_0$ has been locally determined to be $73.02\pm1.79~\rm{km/s/Mpc}$ by using the Wide Field Camera 3 (WFC3) on the Hubble Space Telescope (HST) \citep{Riess:2016jrr}. The $1\sigma$ uncertainty of $H_0$ has been reduced from $3.3\%$ to $2.4\%$. However, this value of local $H_0$ measurement is $3\sigma$ higher than $67.27\pm0.66~\rm{km/s/Mpc}$, which is predicted by the base $\Lambda$CDM model according to the Planck CMB data \citep{Ade:2015xua}. In other words, there is a tension between these two measurements.

The $H_0$ tension might imply some underlying new physics, if it does not arise from some unknown systematic uncertainties. The CMB observations are sensitive to the physics at the last-scattering surface with redshift $z\sim10^3$. By contrast, the local $H_0$ measurement is just sensitive to the late-time physics with redshift $z<0.15$. To resolve the $H_0$ tension, one possible way is to introducing the interaction between cold dark matter and dark energy. The cold dark matter could be converted into the dark energy with the evolution of the Universe. Thus the dark energy will be strengthened in the late-time Universe, and then more efficiently drive the cosmic accelerating expansion. Actually, several papers \citep{Salvatelli:2014zta,Sola:2015wwa,Sola:2016ecz} have provided the first and strong indication of interaction in the dark sector recently.

In this paper, we will study several interacting dark energy (IDE) models by using the latest cosmological observations. Our data compilation include the distance priors, the baryon acoustic oscillation (BAO), the supernovae of type Ia (SNe), the large-scale structure (LSS), the Hubble parameter $H(z)$, and the local $H_0$ measurement. The distance priors were subtracted by \citet{Huang:2015vpa} with the Planck CMB data released in 2015. The BAO data include 6dFGS \citep{Beutler:2011hx}, SDSS MGS \citep{Ross:2014qpa}, and WiggleZ \citep{Kazin:2014qga}. The LSS data include the anisotropic clustering of LOWZ and CMASS galaxies \citep{Gil-Marin:2016wya}. The SNe data refers to the ``Joint Lightcurve Analysis'' (JLA) compilation \citep{Betoule:2014frx}. The $H(z)$ data include 30 data points which are obtained by the differential-age techniques applied to passively evolving galaxies \citep{Zhang:2012mp,Jimenez:2003iv,Simon:2004tf,Stern:2009ep,Moresco:2012jh,Moresco:2016mzx,Moresco:2015cya}. We will use our data combinations to make updated constraints on the interaction between dark sectors. In addition, we will show whether the $H_0$ tension could be reconciled in the framework of IDE models.

The rest of the paper is arranged as follows. In section \ref{idemodels}, we induce the IDE models briefly. In section \ref{method}, the data sets are introduced, together with the method of data analysis. Our constraints on the IDE models are listed in section \ref{result}. Conclusion is given in section \ref{conclusion}.

\section{Interacting dark energy models}\label{idemodels}

We consider the spatially flat Universe in this study. The Friedmann's equation is given by$3M_p^2 H^2 = \rho_{de}+\rho_{c}+\rho_{b}+\rho_{r}$, where $H=d\ln a/dt$ is the Hubble parameter, $M_p = 1/\sqrt{8\pi G}$ denotes the reduced Planck mass, and $\rho_{de}$, $\rho_{c}$, $\rho_{b}$ and $\rho_{r}$ denote the energy densities of dark energy, cold dark matter, baryon, and radiations, respectively. We can define the dimensionless Hubble parameter $E(z)=H(z)/H_0$, which satisfies
\begin{equation}
\label{ez}
E^2 = \Omega_{de0}\frac{\rho_{de}}{\rho_{de0}}+\Omega_{c0}\frac{\rho_{c}}{\rho_{c0}}+\Omega_{b0}\frac{\rho_{b}}{\rho_{b0}}+\Omega_{r0}\frac{\rho_{r}}{\rho_{r0}}\ .
\end{equation}
Here $\Omega_{de0}$, $\Omega_{c0}$, $\Omega_{b0}$ and $\Omega_{r0}$ denote today's energy-density fractions of dark energy, cold dark matter, baryon and radiations, respectively. We have $\rho_{b}=\rho_{b0}(1+z)^3$, and $\rho_{r}=\rho_{r0}(1+z)^4$. Once the equation of state ($w$) of dark energy and the interaction between dark sectors are assumed, $\rho_{de}$ and $\rho_{c}$ can be also expressed in terms of $z$. In addition, we have $\Omega_{r0}=\Omega_{\gamma 0}(1+0.2271N_{eff})$ where $\Omega_{\gamma 0}=2.469\times10^{-5}h^{-2}$, $N_{eff}=3.046$, and $H_0=100h ~\rm{km/s/Mpc}$. Thus the free parameters are $H_0$, $\Omega_{b0}$, $\Omega_{c0}$, $w$, and an interaction parameter. One should note that $\Omega_{de0}$ is a derived parameter, since we have a relation $\Omega_{de0}+\Omega_{m0}+\Omega_{r0}=1$. Here we denote $\Omega_{m0}=\Omega_{c0}+\Omega_{b0}$.

We consider the interaction between dark energy and cold dark matter. The dynamical equations of dark energy and cold dark matter are given by
\begin{eqnarray}
&&\frac{{d\rho}_{de}}{dt}+3H(\rho_{de}+p_{de})=-Q\ ,\\
&&\frac{{\rho}_{c}}{dt}+3H\rho_{c}=Q\ ,
\end{eqnarray}
where $Q$ denotes an interaction term. The above two equations can be rewritten as
\begin{eqnarray}
\label{interactionequation1}
&&(1+z)\frac{d{\rho}_{de}}{dz}-3(1+w)\rho_{de}=\frac{Q}{H}\ ,\\
\label{interactionequation2}
&&(1+z)\frac{d{\rho}_{c}}{dz}-3\rho_{c}=-\frac{Q}{H}\ ,
\end{eqnarray}
where we have used the equation of state of dark energy, i.e. $w=p/\rho$, and noticed relations $z=a^{-1}-1$ and $\frac{d}{dt}=-H(1+z)\frac{d}{dz}$.

The interaction term $Q$ determines the energy transfer rate between dark energy and cold dark matter. However, its specific form is still an open question. One should assume certain possible forms of $Q$ to study the issue of interaction between dark sectors. The following three forms were usually considered, see \citep{Amendola:2006dg,Guo:2007zk,Zhang:2007uh,Costa:2016tpb} and references therein. They are given by
\begin{eqnarray}
&&Q_0=0\ ,\\
&&Q_1=3\gamma H\rho_{de}\ ,\\
&&Q_2=3\gamma H\rho_{c}\ ,
\end{eqnarray}
where $\gamma$ denotes a dimensionless coupling parameter. One should note that the model with $Q_0$ denotes no interaction between dark sectors. Usually, the above three models are denoted by wCDM model, IwCDM1 model and IwCDM2 model, respectively. Particularly, we are interested in some one-parameter generalizations of $\Lambda$CDM model. We will study the IDE models with $w=-1$, which are called I$\Lambda$CDM1 model and I$\Lambda$CDM2 model, respectively.

Once the interaction term $Q$ is determined, one can solve (\ref{interactionequation1}) and (\ref{interactionequation2}) to finally obtain $E(z)$ in (\ref{ez}). For the wCDM model, we deduce $E(z)$ of the form
\begin{equation}
\label{ez1}
E^2(z)=\Omega_{de0}(1+z)^{3(1+w)}+\Omega_{m0}(1+z)^{3}+\Omega_{r0}(1+z)^{4}\ ,
\end{equation}
since there is no interaction between dark sectors. In the case of $w=-1$, we recover the $\Lambda$CDM model. For the IwCDM1 model, we deduce $E(z)$ of the form
\begin{eqnarray}
\label{ez1}
&E^2(z)=\Omega_{de0}\left(\frac{\gamma}{w+\gamma}(1+z)^{3}+\frac{w}{w+\gamma}(1+z)^{3(1+w+\gamma)}\right)\nonumber\\
&+\Omega_{m0}(1+z)^{3}+\Omega_{r0}(1+z)^{4}\ ,
\end{eqnarray}
since (\ref{interactionequation1}) has a solution $\rho_{de}=\rho_{de0}(1+z)^{3(1+w+\gamma)}$.
For the IwCDM2 model, we deduce $E(z)$ of the form
\begin{eqnarray}
\label{ez1}
&E^2(z)=\Omega_{de0}(1+z)^{3(1+w)}+\Omega_{b0}(1+z)^{3}+\Omega_{r0}(1+z)^{4}\nonumber\\
&+\Omega_{c0}\left(\frac{\gamma}{w+\gamma}(1+z)^{3(1+w)}+\frac{w}{w+\gamma}(1+z)^{3(1-\gamma)}\right)\ ,
\end{eqnarray}
since (\ref{interactionequation2}) has a solution $\rho_c=\rho_{c0}(1+z)^{3(1-\gamma)}$. One should let $w=-1$ in the above two expressions, if he wants to study I$\Lambda$CDM1 model and I$\Lambda$CDM2 model.

\section{Data and Methodology}\label{method}

In this study, we will use the latest CMB, BAO and $H_0$ data to constrain the IDE coupling parameter $\gamma$ together with other cosmological parameters. Both the physics of CMB and BAO are well understood, and the systematic uncertainties are under control. Recently, the local value of the Hubble constant $H_0$ has been determined to $2.4\%$ level. However, this value has tension with the prediction of $\Lambda$CDM model which is based on the CMB observations. In this paper, we will show that this tension will disappear in some IDE models. In other words, the $H_0$ data, combined with CMB and BAO, will give a good constraint on the IDE coupling parameter $\gamma$.

For the CMB data, we use the distance priors which are obtained from the Planck data release 2015. One denotes the comoving distance to the last-scattering surface by $r(z_\ast)$, and the comoving sound horizon at the last-scattering epoch by $r_s(z_\ast)$. Then the distance priors are given by these two distance scales through $l_A=\pi r(z_\ast)/r_s(z_\ast)$ and $R=r(z_\ast)\sqrt{\Omega_{m0} H_0^2}$ \citep{Bond:1997wr,Efstathiou:1998xx,Wang:2007mza}, where $z_\ast$ denotes the redshift at the last-scattering surface. Combined with the physical baryon fraction $\omega_b=\Omega_{b0} h^2$, they summarize the CMB data very well. Here the comoving distance to the redshift $z$ is defined by
$r(z)=H_0^{-1}\int_{0}^{z}\frac{dz^\prime}{E(z^\prime)}$, for the spatially flat Universe. The comoving sound horizon to the last-scattering surface is given by
$r_s(z_\ast)=H_0^{-1}\int_{0}^{\frac{1}{1+z_\ast}}\frac{da}{a^2E(a)\sqrt{3\left(1+{3\Omega_{b0}}/{(4\Omega_{\gamma 0})}a\right)}}$. Here the fitting formula of $z_\ast$ is given by
$z_\ast=1048\left(1+0.00124\omega_{b}^{-0.738}\right)\left(1+g_1\omega_{b}^{g_{2}}\right)$ \citep{Hu:1995en}, where we have $g_1=0.0783\omega_{b}^{-0.238}/(1+39.5\omega_{b}^{0.763})$ and $g_2=0.560/(1+21.1\omega_{b}^{1.81})$.
By using the Planck TT,TE,EE+lowP data, recently, the distance priors were subtracted by \citet{Huang:2015vpa}. They are listed in Table~\ref{tab:distancepriors}, together with their normalized covariance matrix $NormCov_{CMB}(p_i,p_j)$ where $i=1,2,3$.
\begin{table}
\centering
\renewcommand{\arraystretch}{1.5}
\begin{tabular}{ccccc}
\hline\hline
    & $Planck~ \textrm{TT,TE,EE}+\textrm{lowP}$ & $R$ & $l_\textrm{A}$ & $\omega_b$\\
\hline
$R$                                           & $1.7448\pm0.0054$       & $1.0$&    $0.53$&   $-0.73$\\
$l_\textrm{A}$                                & $301.460\pm0.094$        & $0.53$&    $1.0$&   $-0.42$ \\
$\omega_b$                              & $0.02240\pm0.00017$      & $-0.73$&   $-0.42$&  $1.0$ \\
\hline\hline
\end{tabular}
\caption{Distance priors from the Planck TT,TE,EE+lowP data, together with their normalized covariance matrix \citep{Huang:2015vpa}.}
\label{tab:distancepriors}
\end{table}
The covariance matrix can be obtained via $Cov_{CMB}(p_i,p_j)=\sigma(p_i)\sigma(p_j)NormCov_{CMB}(p_i,p_j)$. The $\chi_{CMB}^2$ of the distance priors is given by
$\chi_{CMB}^2=(p_i-p_i^{obs})Cov^{-1}_{CMB}(p_i,p_j)(p_j-p_j^{obs})$,
where $p$ and $p^{obs}$ denote the theoretical values and the observational mean values, respectively. Here we also listed the base parameter $\omega_b$.

For the BAO data, we use the isotropic BAO estimator $r_s(z_d)/D_V(z)$ of 6dFGS at an effective redshift $z_{6dFGS}=0.106$ \citep{Beutler:2011hx} and SDSS MGS at $z_{MGS}=0.15$ \citep{Ross:2014qpa}, and WiggleZ at $z_{WiggleZ}=0.44, 0.6, 0.73$ \citep{Kazin:2014qga}. We take into account the correlations among the WiggleZ data points. Here $r_s(z_d)$ denotes the comoving sound horizon at the baryon-drag epoch $z_d$, and $D_V(z)=\left[(1+z)^2D_A^2(z)cz/H(z)\right]^{1/3}$ where $D_A(z)=r(z)/(1+z)$ is the angular diameter distance. The BAO distance measurements can help to break the geometric degeneracy. The $\chi^2$ of the BAO data is denoted by $\chi^2_{BAO}$.

For the LSS data, we refer to the anisotropic clustering of LOWZ and CMASS galaxies \citep{Gil-Marin:2016wya}, which contain the geometric information from the Alcock-Paczynski (AP) effect \citep{Alcock:1979mp}. The LOWZ sample is located at an effective redshift $z_{\rm{LOWZ}}=0.32$, and the CMASS sample at $z_{\rm{CMASS}}=0.57$. The AP effect is sensitive to $F_{AP}(z)=(1+z)D_A(z)H(z)$. The anisotropic BAO estimators are given by $D_A(z)/r_s(z_d)$ and $H(z)r_s(z_d)$. They contain the information of $D_V/r_s(z_d)$ and $F_{AP}$ simultaneously. We will use the data of $D_A/r_s(z_d)$ and $Hr_s(z_d)$ together with their covariance matrix. The LSS data may further break the geometric degeneracy. Here the $\chi^2$ of the LSS data is denoted by $\chi^2_{LSS}$.

For the SNe data, we use the JLA compilation \citep{Betoule:2014frx}. Theoretically, the luminosity distance at redshift $z$ is given by $D_L(z)=(1+z)r(z)$. For the JLA, the luminosity distance of a supernova is $D_L(z_{hel},z_{cmb})=(1+z_{hel})r(z_{cmb})$, where $z_{cmb}$ and $z_{hel}$ denote the CMB frame redshift and the heliocentric redshift, respectively. The distance modulus is defined as $\mu=5\log_{10}D_L/10\textrm{pc}$. The $\chi^2_{SNe}$ of the JLA SNe is given by $\chi^2_{SNe}=(\mu_{obs}-\mu_{th})^\dagger C^{-1}(\mu_{obs}-\mu_{th})$, where $C$ is a covariance matrix.

For the $H(z)$ data, we use 30 data points listed in Table~\ref{tab:hz}. They are obtained by the differential-age techniques applied to passively evolving galaxies \citep{Zhang:2012mp,Jimenez:2003iv,Simon:2004tf,Stern:2009ep,Moresco:2012jh,Moresco:2016mzx,Moresco:2015cya}, and then there are no correlations with the BAO data. The $\chi^2_{H(z)}$ of the $H(z)$ data is given by $\chi^2_{H(z)}=\left(\frac{H(z)-H^{obs}(z)}{\sigma_{H(z)}}\right)^2$, where $H(z)$ is the theoretical Hubble parameter, $H^{obs}(z)$ and $\sigma_{H(z)}$ are the observed Hubble parameter and its $1\sigma$ uncertainty, respectively.
\begin{table}
\centering
\renewcommand{\arraystretch}{1.5}
\scriptsize
\begin{tabular}{ccc}
\hline\hline
 $z$ & $H(z)$ & \textrm{Ref.} \\
\hline
$0.07$   & $69.0\pm19.6$   &  \citet{Zhang:2012mp} \\
$0.09$    & $69.0\pm12.0$   &  \citet{Jimenez:2003iv} \\
$0.12$   & $68.6\pm26.2$   &  \citet{Zhang:2012mp} \\
$0.17$   & $83.0\pm8.0$   &  \citet{Simon:2004tf} \\
$0.1791$   & $75.0\pm4.0$   &  \citet{Moresco:2012jh} \\
$0.1993$   & $75.0\pm5.0$   &  \citet{Moresco:2012jh} \\
$0.2$   & $72.9\pm29.6$   &  \citet{Zhang:2012mp} \\
$0.27$   & $77.0\pm14.0$   &  \citet{Simon:2004tf} \\
$0.28$   & $88.8\pm36.6$   & \citet{Zhang:2012mp}  \\
$0.3519$   & $83.0\pm14.0$   &  \citet{Moresco:2012jh} \\
$0.3802$   & $83.0\pm13.5$   &  \citet{Moresco:2016mzx} \\
$0.4$   & $95.0\pm17.0$   &  \citet{Simon:2004tf} \\
$0.4004$   & $77.0\pm10.2$   &  \citet{Moresco:2016mzx} \\
$0.4247$   & $87.1\pm11.2$   &  \citet{Moresco:2016mzx} \\
$0.4497$   & $92.8\pm12.9$   &  \citet{Moresco:2016mzx} \\
$0.4783$   & $80.9\pm9.0$   & \citet{Moresco:2016mzx}  \\
$0.48$   & $97.0\pm62.0$   & \citet{Stern:2009ep}  \\
$0.5929$   & $104.0\pm13.0$   &  \citet{Moresco:2012jh} \\
$0.6797$   & $92.0\pm8.0$   &  \citet{Moresco:2012jh} \\
$0.7812$   & $105.0\pm12.0$   &  \citet{Moresco:2012jh} \\
$0.8754$   & $125.0\pm17.0$   &  \citet{Moresco:2012jh} \\
$0.88$   & $90.0\pm40.0$   &  \citet{Stern:2009ep} \\
$0.9$   & $117.0\pm23.0$   &  \citet{Simon:2004tf} \\
$1.037$   & $154.0\pm20.0$   &  \citet{Moresco:2012jh} \\
$1.3$   & $168.0\pm17.0$   &  \citet{Simon:2004tf} \\
$1.363$   & $160.0\pm33.6$   &  \citet{Moresco:2015cya} \\
$1.43$   & $177.0\pm18.0$   &  \citet{Simon:2004tf} \\
$1.53$   & $140.0\pm14.0$   &  \citet{Simon:2004tf} \\
$1.75$   & $202.0\pm40.0$   &  \citet{Simon:2004tf} \\
$1.965$   & $186.5\pm50.4$   &  \citet{Moresco:2015cya} \\
\hline\hline
\end{tabular}
\caption{The data list of the observed Hubble parameters $H(z)$ [$\rm{km~s^{-1}~Mpc^{-1}}$].}
\label{tab:hz}
\end{table}

Recently the uncertainty of the local value of the Hubble constant has been reduced from $3.3\%$ to $2.4\%$ by using the WFC3 on the HST. The best estimate of $H_0$ is given by \citep{Riess:2016jrr}
\begin{equation}
H_0^{obs}=73.02\pm1.79~\rm{km~s^{-1}~Mpc^{-1}}
\end{equation}
at $1\sigma$ confidence level. This value is in tension with the $\Lambda$CDM prediction which is based on the CMB observations. For example, it is $3.0\sigma$ higher than the $67.27\pm0.66~\rm{km~s^{-1}~Mpc^{-1}}$ which is predicted by the base $\Lambda$CDM model and Planck CMB data \citep{Ade:2015xua}. In this study, we try to resolve this tension in the framework of IDE models. The $\chi^2_{H_0}$ of the local $H_0$ data is given by
$\chi^2_{H_0}=\left(\frac{H_0-H_0^{obs}}{\sigma_{H_0}}\right)^2$, where $\sigma_{H_0}$ denotes the $1\sigma$ uncertainty of local $H_0$, and $H_0^{obs}$ is the mean value of local $H_0$. The distance priors are sensitive to the physics with the redshift $z\sim10^3$. By contrast, the $H_0$ observation is corresponded to the late-time physics $z<0.15$. In other words, a higher value of local $H_0$ may reveal that the dark energy is strengthened in the late-time Universe.

We employ the Cosmological Monte Carlo (CosmoMC) sampler \citep{Lewis:2002ah} to estimate the parameter space of the IDE models. The Gelman and Rubin criterion is set by $R-1=0.01$ to ensure the statistical convergence. We use the data combination CMB+BAO+SNe+LSS+$H(z)$+$H_0$ in this study. The joint likelihood is given by $\mathcal{L}\propto e^{-\chi^2/2}$, where $\chi^2=\chi^2_{CMB}+\chi^2_{BAO}+\chi^2_{SNe}+\chi^2_{LSS}+\chi^2_{H(z)}+\chi^2_{H_0}$.
For the wCDM model, the parameter space is spanned by $\{\Omega_{c0},\Omega_{b0},H_0,w\}$. For the IwCDM models, the parameter space is spanned by $\{\Omega_{c0},\Omega_{b0},H_0,w,\gamma\}$. Here $\Omega_{de0}$ is a derived parameter. In addition, we also consider the I$\Lambda$CDM models for which $w=-1$, and the parameter space spanned by $\{\Omega_{c0},\Omega_{b0},H_0,\gamma\}$. One should note that $\Omega_{de0}$ is a derived parameter.

We employ the Deviance Information Criterion (DIC) to judge either a model $M_1$ or a model $M_2$ is preferred by a given data set $D$. To describe the goodness of fit, as mentioned above, we calculate $\chi^2(p)=-2\ln \mathcal{L}(D|p,M_i)$ where $p$ denotes a set of parameters of the model $M_i$. The mean goodness of fit is given by $\langle\chi^2\rangle=-2\langle\ln \mathcal{L}\rangle$. \citet{Spiegelhalter001} define the DIC as
$DIC(M_i)=\langle\chi^2\rangle+p_D$, where $p_D$ denotes the Bayesian complexity describing the effective complexity of the model. The Bayesian complexity is defined by
$p_D=\langle\chi^2\rangle-\chi^2(\tilde{p})$, where $\tilde{p}$ is the maximum likelihood point in the parameter space. A lower DIC implies either the model fits the data better (a lower $\langle\chi^2\rangle$) or the model has less complexity. We refer to the difference between the DICs of two models, namely, $\Delta DIC=DIC(M_1)-DIC(M_2)$. If $\Delta DIC=0$, neither model is preferred by the data. If $0<\Delta DIC<2$, the data indicates no significant preference for $M_2$. If $2<\Delta DIC<6$, there is a positive preference for $M_2$. If $\Delta DIC>6$, the preference is strong. By contrast, the negative values mean that the data prefers $M_1$.

\section{Results}\label{result}

Our constraints on cosmological parameters are summarized in Table~\ref{tab:ilcdm} for the $\Lambda$CDM model and two I$\Lambda$CDM models.
\begin{table}
\centering
\renewcommand{\arraystretch}{1.5}
\begin{tabular}{cccc}
\hline\hline
                & $\Lambda$CDM             & I$\Lambda$CDM1           & I$\Lambda$CDM2 \\
\hline
$\Omega_{b0}$   & $0.0476\pm0.0005$        &  $0.0472\pm0.0007$       &   $0.0475\pm0.0007$ \\
$\Omega_{c0}$   & $0.2520\pm0.0058$        &  $0.2502\pm0.0059$       &   $0.2528\pm0.0065$ \\
$\Omega_{de0}$  & $0.7003\pm0.0063$        &  $0.7026\pm0.0065$       &   $0.6997\pm0.0067$ \\
$H_0$           & $68.75\pm0.49$           &  $68.97\pm0.54$          &   $68.90\pm0.64$ \\
$\gamma$        &  $-$           &  $0.0027\pm0.0037$       &   $-0.0004\pm0.0015$ \\
\hline
$\chi^2_{\rm{min}}$        & $724.07$        &  $723.42$           &   $724.04$ \\
DIC        & $734.97$                   &  $737.02$                &   $738.45$ \\
\hline\hline
\end{tabular}
\caption{Constraints on the free parameters of $\Lambda$CDM, I$\Lambda$CDM1, and I$\Lambda$CDM2 models. The derived parameter $\Omega_{de0}$ and the best-fit $\chi^2$ are also listed here, as well as the DIC. The dimension of $H_0$ is $\rm{km~s^{-1}~Mpc^{-1}}$.}
\label{tab:ilcdm}
\end{table}
For the $\Lambda$CDM model, the best-fit value of $H_0$, i.e. $H_0=68.75\pm0.49~\rm{km/s/Mpc}$, is much lower than the local value of $H_0$ by $2.4\sigma$. Similar situations are showed for both I$\Lambda$CDM models. The dimensionless coupling parameter $\gamma$ is consistent with zero for both I$\Lambda$CDM models. By contrast to the $\Lambda$CDM model, the data combination prefer neither the I$\Lambda$CDM1 model nor the I$\Lambda$CDM2 model. For both I$\Lambda$CDM models, the minimum $\chi^2$ are similar to that of the $\Lambda$CDM model, but their DIC are larger than that of the $\Lambda$CDM model. The data combination show a preference for the $\Lambda$CDM model.

Our constraints on cosmological parameters are summarized in Table~\ref{tab:iwcdm} for the wCDM model and two IwCDM models.
\begin{table}
\centering
\renewcommand{\arraystretch}{1.5}
\begin{tabular}{cccc}
\hline\hline
                & wCDM                     & IwCDM1                   & IwCDM2 \\
\hline
$\Omega_{b0}$   & $0.0459\pm0.0013$        &  $0.0459\pm0.0013$       &   $0.0450\pm0.0016$ \\
$\Omega_{c0}$   & $0.2466\pm0.0065$        &  $0.2472\pm0.0066$       &   $0.2486\pm0.0069$ \\
$\Omega_{de0}$  & $0.7075\pm0.0075$        &  $0.7069\pm0.0077$       &   $0.7063\pm0.0077$ \\
$H_0$           & $69.88\pm0.90$           &  $69.87\pm0.98$          &   $70.65\pm1.23$ \\
$w$             & $-1.055\pm0.039$         &  $-1.064\pm0.053$          &   $-1.071\pm0.043$ \\
$\gamma$        &  $-$    &  $-0.0014\pm0.0051$      &   $-0.0015\pm0.0016$ \\
\hline
$\chi^2_{\rm{min}}$        & $722.19$       &  $722.36$           &   $721.33$ \\
DIC        & $734.04$                   &  $736.59$                  &   $735.33$ \\
\hline\hline
\end{tabular}
\caption{Constraints on the free parameters of wCDM, IwCDM1, and IwCDM2 models. The derived parameter $\Omega_{de0}$ and the best-fit $\chi^2$ are also listed here, as well as the DIC. The dimension of $H_0$ is $\rm{km~s^{-1}~Mpc^{-1}}$.}
\label{tab:iwcdm}
\end{table}
The data combination prefers $w<-1$ at the $1.4\sigma$ level, namely, we have $w=-1.055\pm0.039$. However, the best-fit value of $H_0$, i.e. $H_0=69.88\pm0.90~\rm{km/s/Mpc}$, is still lower than the local value of $H_0$ by $1.75\sigma$. By contrast to the $\Lambda$CDM model, the $H_0$ tension is slightly alleviated in the wCDM model, but not enough. Based on $\Delta DIC=DIC_{wCDM}-DIC_{\Lambda CDM}=-0.93$, we find that there is no significant preference for the wCDM model. In addition, the wCDM model fits the data better than the $\Lambda$CDM model, since the minimum $\chi^2$ is reduced by $1.88$.

By contrast to the wCDM model, the $H_0$ tension is still remained in the IwCDM1 model, even though we consider the interaction effect between the dark sector. We obtain $H_0=69.87\pm0.98~\rm{km/s/Mpc}$ which is lower than the local $H_0$ measurement by $1.76\sigma$. The best-fit value of $w$, i.e. $w=-1.064\pm0.053$, is also smaller than $-1$ at the $1.2\sigma$ level. The dimensionless coupling parameter, i.e. $\gamma=-0.0014\pm0.0051$, is consistent with zero. In this model, the minimum $\chi^2$ is is smaller by $1.71$ than that of the $\Lambda$CDM model. However, the DIC becomes larger by $1.62$. Thus this model is not significantly preferred by the data, even though it fits the data better.

For the IwCDM2 model, the best-fit value of $H_0$, i.e. $H_0=70.65\pm1.23~\rm{km/s/Mpc}$, is lower than the local $H_0$ measurement by $1.3\sigma$. The $H_0$ tension is moderately alleviated in this model. The best-fit value of $w$, i.e. $w=-1.071\pm0.043$, is smaller than $-1$ by around $1.7\sigma$. The dimensionless coupling parameter, i.e. $\gamma=-0.0015\pm0.0016$, which is consistent with zero within $1\sigma$. By contrast to the $\Lambda$CDM model, the $\chi^2$ for the IwCDM2 model becomes smaller by $2.74$. Since the DIC becomes larger by $0.36$, there is no significant preference for the IwCDM2 model.

\begin{figure}
 \includegraphics[width=3in]{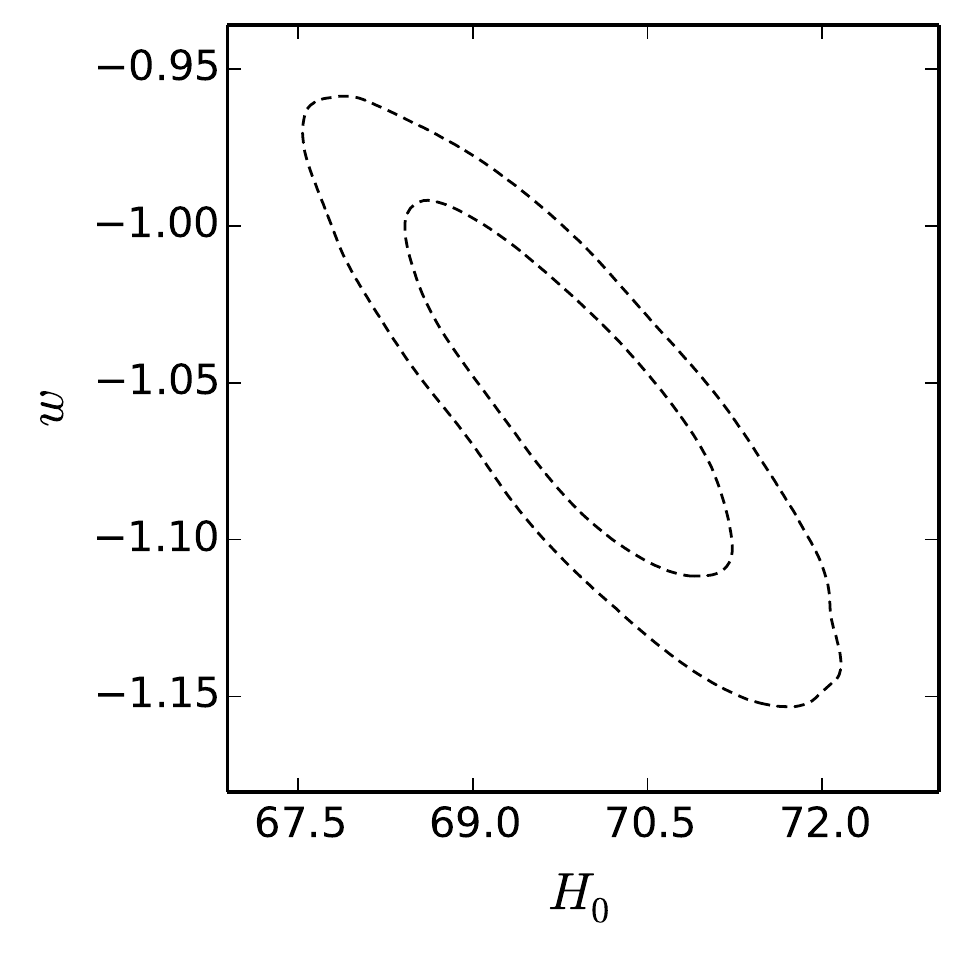}
 \caption{The marginalized distribution contour of $H_0$ and $w$ for the wCDM model.}
 \label{fig:wcdm}
\end{figure}

\begin{figure}
 \includegraphics[width=3.36in]{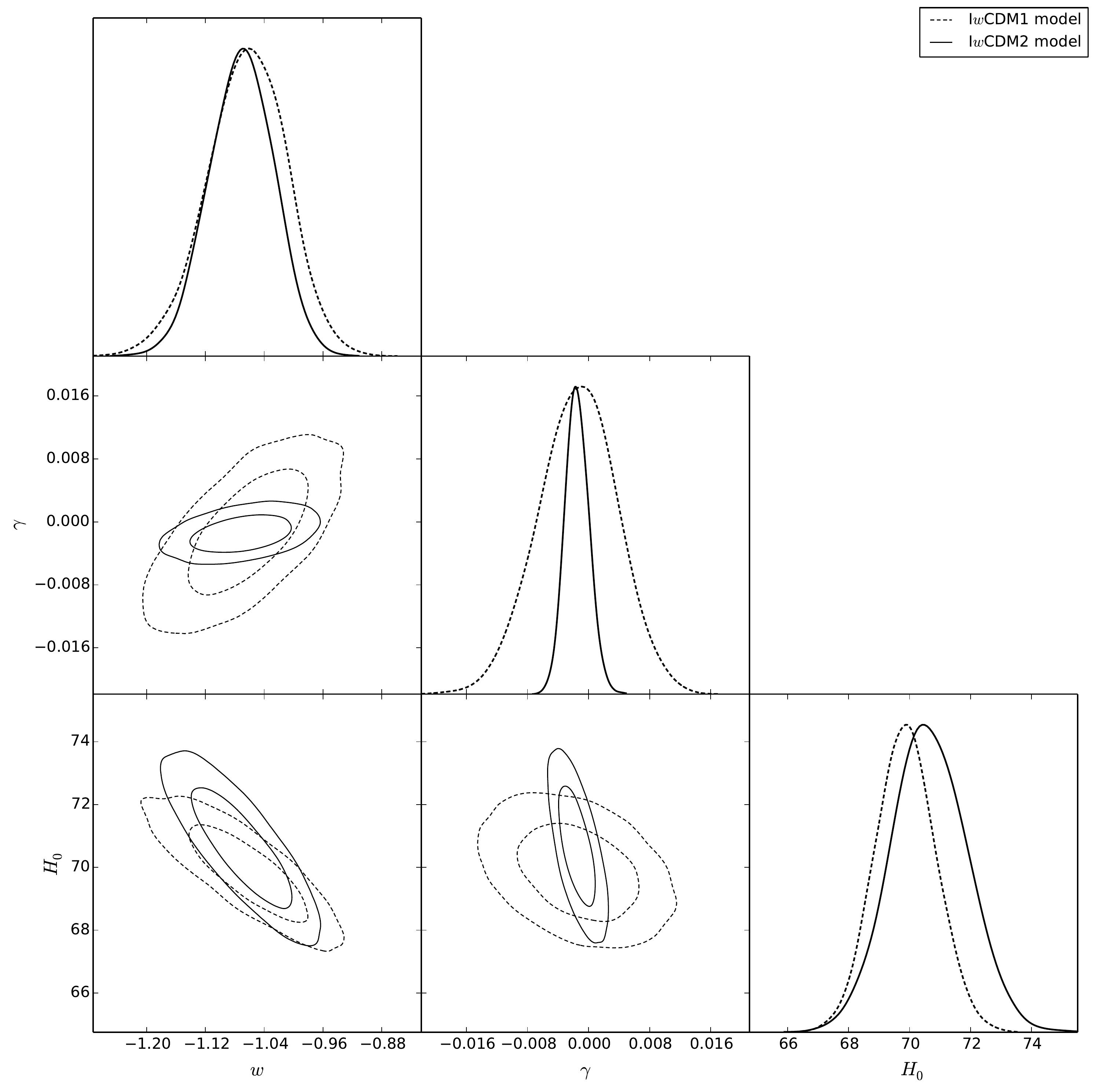}
 \caption{The marginalized distribution contours and the likelihood distributions of $w$, $\gamma$ and $H_0$ for the IwCDM models.}
 \label{fig:tri}
\end{figure}

To directly show how the wCDM model alleviates the $H_0$ tension, we plot the marginalized distribution contour of $H_0$ and $w$ in Figure~\ref{fig:wcdm}.
We find that $H_0$ is strongly anti-correlated with $w$ in the $H_0$-$w$ plane. Thus a higher value of $H_0$ can be accounted by a smaller value of $w$. To reveal how the local $H_0$ data constrains the IwCDM models, we plot the marginalized distribution contours and the likelihood distributions of $H_0$, $w$, and $\gamma$ in Figure~\ref{fig:tri}.
Similar to the wCDM model, $H_0$ is also anti-correlated with $w$ in both IwCDM models. It is further anti-correlated with $\gamma$. This means that a higher value of $H_0$ requires more energy density flowing from cold dark matter to dark energy. Unfortunately, both IwCDM models can not totally resolve the $H_0$ tension, but just alleviate.

Our above results can be compared with recent results obtained by other authors. For instance, \citet{Costa:2016tpb} made updated constraints for IwCDM1 and IwCDM2 by using the Planck+BAO+SNIa+RSD+$H_0$ data. In this paragraph, Planck denotes Planck 2015 CMB data instead of the distance priors; BAO denotes the isotropic 6dFGS, MGS, BOSS DR11 LOWZ and CMASS; the value of $H_0$ is lower than the report of \citet{Riess:2016jrr}. The authors found the interaction between dark sectors strongly suppressed. This is compatible with our result in this study. \citet{Murgia:2016ccp} made updated constraints for the IwCDM1 model with two sets of priors of parameters by using the Planck+BAO+SNIa+RSD+gravitational lensing data. By assuming that dark matter decays into dark energy, the tension with the independent determinations of $H_0$ and $\sigma$ increases. When dark matter is fed by dark energy, the tension can be nicely released. \citet{Nunes:2016dlj} made updated constraints for the IwCDM2 model by using the Cosmic chronometers+Planck+BAO+SNIa+$H_0$ data. Here the $H_0$ data comes from  \citet{Riess:2016jrr}. The authors found that the direct between interaction between dark sectors is mildly favored, while the EoS of dark energy is $w<-1$ at the $3\sigma$ level. This is different from our result that $w<-1$ at the $1.7\sigma$ level.

\section{Conclusion}\label{conclusion}

The cosmological observations have provided us highly precise data. Recently, the local measurement showed a higher value of $H_0$ than the prediction of $\Lambda$CDM model based on the CMB data. This fact might reveal either some tensions exist between the local $H_0$ measurement and the CMB observations, or there is underlying new physics. For example, dark energy may be strengthened in the late-time Universe. In this paper, we explored several IDE models with the latest cosmological observations including the data of CMB, BAO, LSS, SNe, $H(z)$ and $H_0$. In the IDE models, the interaction between dark sectors may strengthen dark energy. This fact could help to reconcile the $H_0$ tension.

Our results showed that the local value of $H_0$ is still in tension with two I$\Lambda$CDM models considered in this study. However, the wCDM model can slightly alleviate this tension. The higher value of local $H_0$ implies a more negative value of $w$. We obtained $w=-1.055\pm0.039$ in this case. The interaction between dark sectors could further release the $H_0$ tension. The data combination provided severe constraints on the interacting coupling parameter $\gamma$ in two IwCDM models. We obtained $w=-1.064\pm0.053$ and $\gamma=-0.0014\pm0.0051$ for the IwCDM1 model, and $w=-1.071\pm0.043$ and $\gamma=-0.0015\pm0.0016$ for the IwCDM2 model. Therefore, we found no significant preference for the interaction between dark energy and dark matter.

\section*{Acknowledgments}

The author (DMX) is supported by National Natural Science Foundation of China (NSFC) (grant NO. 11505018), and Chongqing Science and Technology Plan Project (grant NO. Cstc2015jvyj40031). We thank Ke~Wang for helpful discussions when we are preparing this paper.


\end{document}